\documentclass[options]{JHEP3}
\title{Non-Gaussianity and finite length inflation}
\author{Shiro Hirai\\ 
Department of Digital Games, Osaka Electro-Communication University\\
1130-70 Kiyotaki, Shijonawate, Osaka 575-0063, Japan\\ 
E-mail: \email{hirai@isc.osakac.ac.jp}}
\author{Tomoyuki Takami\\ 
Department of Digital Games, Osaka Electro-Communication University\\
1130-70 Kiyotaki, Shijonawate, Osaka 575-0063, Japan\\
E-mail: \email{takami@isc.osakac.ac.jp}}
\abstract{In the present paper, certain inflation models are shown to have large non-Gaussianity in special cases. Namely, finite length inflation models with an effective higher derivative interaction, in which slow-roll inflation is adopted as inflation and a scalar-matter-dominated period or power inflation is adopted as pre-inflation, are considered. Using Holman and Tolley's formula of the nonlinearity parameter $f^\textrm{\tiny flattened}_\textrm{\tiny NL}$, we calculate the value of $f^\textrm{\tiny flattened}_\textrm{\tiny NL}$. A large value of $f^\textrm{\tiny flattened}_\textrm{\tiny NL}(f^\textrm{\tiny flattened}_\textrm{\tiny NL} > 100)$ can be obtained for all of the models considered herein when the length of inflation is 60-63 $e$-folds and $f_\textrm{\tiny NL}$ has strong dependence on the length of inflation. Interestingly, this length is similar to that for the case in which the suppression of the CMB angular power spectrum of $l=2$ was derived using the inflation models described in our previous papers.}

\preprint{\today}
\dedicated{...}
\usepackage[dvips]{graphicx}

\begin{document}
\section{Introduction}
Non-Gaussianity of primordial perturbations is one of the most interesting problems implied by the WMAP data [1, 2]. The observational limits on the nonlinearity parameter from WMAP seven-year data [2] are $-10<f^\textrm{\tiny local}_\textrm{\tiny NL}<74$ (95\% CL) , $-214<f^\textrm{\tiny equil}_\textrm{\tiny NL}<266$ (95\% CL) and $-410<f^\textrm{\tiny orthog}_\textrm{\tiny NL}<6$ (95\% CL). 
However, the standard simple inflation model predicts approximately Gaussian fluctuation, the deviation from Gaussian of which is very small. Several studies have attempted to achieve such large non-Gaussianity. Holman and Tolley [3] showed that if the effective action for the inflaton contains a higher-derivative interaction, which is derived, for example, from k-inflation [4] or DBI inflation [5], and the initial state of inflation is not the Bunch-Davies vacuum, then enhanced non-Gaussianity is derived in the "flattened" triangle configurations, the contribution of which is also discussed in [6].  In their paper, the initial state of the curvature perturbation in inflation was assumed not to be the Bunch-Davies vacuum, i.e., squeezed states, but they did not report a concrete value or the physical mechanism that generates the initial state in inflation, although the value of the coefficient of the initial state in inflation has a very important effect on the non-Gaussianity.

On the other hand, the effect of the initial condition in inflation on the power spectrum of curvature perturbations has been considered [7-9] and the effect of the length of inflation and pre-inflation physics on the power spectrum and the angular power spectrum of scalar and tensor perturbations has been examined by the present authors. [8-9]. The suppression of the spectrum at $l=2$  as indicated by Wilkinson Microwave Anisotropy Probe (WMAP) data [1] may be explained to a certain extent by the finite length of inflation for an inflation of 50-60 $e$-folds [9]. Of course, there are many attempts [10] to derive this suppression. Based on the physical conditions before inflation, we have shown that the initial state of scalar perturbations in inflation is not simply the Bunch-Davies state, but rather a more general state (a squeezed state), where a scalar-matter-dominated period, a radiation-dominated period, or another inflation is considered as pre-inflation, and the general initial states in inflation were calculated analytically. In the present paper, we demonstrate a new property of the proposed inflation model. Using Holman and Talley's formula for the nonlinearity parameter $f^\textrm{\tiny flattened}_\textrm{\tiny NL}$, we calculate the value of $f^\textrm{\tiny flattened}_\textrm{\tiny NL}$ for the case in which the proposed finite inflation models have effective higher-derivative interactions, where slow-roll inflation is adopted as inflation and a scalar-matter-dominated period or power-law inflation period is adopted as pre-inflation. The obtained results are very interesting.
\section{Scalar perturbations}

We consider curvature perturbations in inflation and a scalar-matter-dominated epoch. The background spectrum considered is a spatially flat Friedman-Robertson-Walker (FRW) universe described by metric perturbations. The line element for the background and perturbations is generally expressed as [11] 
\begin{equation}
ds^2 = a^2(\eta)\{(1+2A)d\eta^2-2\partial_i Bdx^i d\eta -[(1-2\Psi)\delta_{ij}+2\partial_i \partial_j E+h_{ij}]dx^i dx^j\},
\end{equation}
where $\eta$ is the conformal time, the functions $A$, $B$, $\Psi$, and $E$ represent the scalar perturbations, and $h_{ij}$ represents tensor perturbations. The density perturbation in terms of the intrinsic curvature perturbation of comoving hypersurfaces is given by $\mathcal{R}=-\Psi-(H/\dot\phi)\delta\phi$, where $\phi$ is the inflaton field, $\delta\phi$ is the fluctuation of the inflaton field, $H$ is the Hubble expansion parameter, and $\mathcal{R}$ is the curvature perturbation. Overdots represent derivatives with respect to time $t$, and primes represent derivatives with respect to the conformal time $\eta$. Introducing the gauge-invariant potential 
$u \equiv  a(\eta)(\delta\phi+(\dot\phi/H)\Psi)$ allows the action for scalar perturbations to be written as [12]
\begin{equation}
S = \frac{1}{2}\int d\eta d^3x \{(\frac{\partial u}{\partial \eta})^2-c^2_{\small \textrm{s}}(\nabla u)^2+\frac{Z''}{Z}u^2\},
\end{equation}
where $c_{\small \textrm{s}}$ is the velocity of sound, and in inflation $Z=a\dot\phi/H$, and $u=-Z\mathcal{R}$. The field $u(\eta, \mathbf{x})$ is expressed using annihilation and creation operators as follows:
\begin{equation}
u(\eta, \mathbf{x}) =\frac{1}{(2\pi)^{3/2}}\int d^3k {\{u_k(\eta)\mathbf{a}_\mathbf{k}+u^{*}_k(\eta)\mathbf{a}^{\dagger}_{-\mathbf{k}}}\}e^{-i\mathbf{kx}}.
\end{equation}
The field equation for $u_k(\eta)$ is derived as
\begin{equation}
\frac{d^2u_k}{d\eta^2}+(c^2_{\small \textrm{s}} k^2-\frac{1}{Z}\frac{d^2Z}{d\eta^2})u_k=0,
\end{equation}
where $c^2_{\small \textrm{s}}=1$ is assumed in inflation. The solution of $u_k$ satisfies the normalization condition $u_kdu^{*}_k/d\eta-u^{*}_k du_k/d\eta=i$. 

First, slow-roll inflation is considered. The slow-roll parameters are defined as [13,14]:
\begin{equation}
\epsilon=3\frac{\dot{\phi }^2}{2}(\frac{\dot{\phi }^2}{2}+V)^{-1}  =
2M^2_\mathrm{P} (\frac{H'(\phi )}{H(\phi )})^2,
\end{equation}
\begin{equation}
\delta=2M^2_\mathrm{P}\frac{H''(\phi)}{H(\phi)},
\end{equation}
\begin{equation}
\xi=4M^{4}_\mathrm{P}\frac{H'(\phi)H'''(\phi)}{(H(\phi))^2}.
\end{equation}
The quantity $V(\phi)$ is the inflation potential, and $M_\mathrm{P}$ is the reduced Plank mass. Other slow-roll parameters ($\epsilon_\textrm{\tiny V}, \eta_\textrm{\tiny V}, \xi_\textrm{\tiny V}$) can be written in terms of the slow-roll parameters $\epsilon$, $\delta$, and $\xi$ for  first-order slow roll, i.e., $\epsilon=\epsilon_\textrm{\tiny V}$, $\delta=\eta_\textrm{\tiny V}-\epsilon_\textrm{\tiny V}$, and $\xi=\xi_\textrm{\tiny V}-3\epsilon_\textrm{\tiny V}\eta_\textrm{\tiny V} + 3\epsilon^2_\textrm{\tiny V}$, where $\epsilon_\textrm{\tiny V}=M^2_\mathrm{P}/2(V'/V)^2$, $\eta_\textrm{\tiny V}=M^2_\mathrm{P}(V''/V^2)$, and $\xi_\textrm{\tiny V}=M^2_\mathrm{P} (V'V'''/V^2)$. Using the slow-roll parameters, $(d^2Z/d\eta^2)/Z$ is written exactly as
\begin{equation}
\frac{1}{Z}\frac{d^2 Z}{d\eta^2}=2a^2 H^2 (1+\epsilon-\frac{3}{2}\delta+\epsilon^2-2\epsilon\delta+\frac{\delta^2}{2}+\frac{\xi}{2}),
\end{equation}
and the scale factor is written as $a(\eta)=-((1-\epsilon)\eta H)^{-1}$. Here, the slow-roll parameters are assumed to satisfy $\epsilon<1$, $\delta<1$, and $\xi<1$. As only the leading-order terms of $\epsilon$ and $\delta$ are adopted, $\epsilon$ and $\delta$ may be considered to be constant, allowing the scale factor to be written as $a(\eta)\approx (-\eta)^{-1-\epsilon}$[14]. Equation (2.4) can be rewritten as 
\begin{equation}
\frac{d^2 u_{k}}{d\eta^2}+(k^2-\frac{2+6\epsilon-3\delta}{\eta^2})u_k=0.
\end{equation}
The solution to Eq. (2.9) is written as [13]
\begin{equation}
f^\textrm{\tiny I}_k(\eta)=\frac{\sqrt{\pi}}{2}e^{i(\nu+1/2)\pi/2}(-\eta)^{1/2}H^{(1)}_{\nu}(-k\eta),
\end{equation}
where $\nu=3/2+2\epsilon-\delta$, and $H^{(1)}_{\nu}$ is the Hankel function of the first kind of order $\nu$. The mode functions $u_k(\eta)$ of a general initial state in inflation are written as 
\begin{equation}
u_k(\eta)=c_1 f^\textrm{\tiny I}_{k}(\eta)+c_2 f^\textrm{\tiny I*}_{k}(\eta),
\end{equation}
where the coefficients $c_1$ and $c_2$ obey the relation $|c_1|^2-|c_2|^2=1$. The important point here is that the coefficients $c_1$ and $c_2$ do not change during inflation. In ordinary cases, the field $u_k(\eta)$ is considered to be in the Bunch-Davies state, i.e., $c_1=1$ and $c_2=0$, because as $\eta\to-\infty$, the field $u_k(\eta)$ must approach plane waves ($e^{-ik\eta}/\sqrt{2k}$).  Second, in the case of power-law inflation, where $a(t)\propto t^q$, a similar method can be used, and the solution is obtained as Eq.(2.10) with $\nu=3/2+1/(q-1)$. Third, the curvature perturbations in the scalar matter are calculated using a method similar to that used for inflation [7, 12, 15].  The field equation $u_k$ can be written in a form similar to Eq. (2.4) with a value of $c^2_{\small \textrm{s}}=1$ and with $Z\propto a_\mathrm{p}(\eta)[\mathcal{H}^2-\mathcal{H}']^{1/2}/\mathcal{H}$, (where $\mathcal{H} =a'_\mathrm{p}/a_\mathrm{p}$). The solution to Eq. (2.4) is then written as $f^\textrm{\tiny S}_{k}=(1-i/(k\eta)\textrm{exp}[-ik\eta])/\sqrt{2k}$.

\section{Calculation of the nonlinearity parameter}

Here, an inflation model is considered.  Since we consider slow-roll inflation to have a finite length, we assume a pre-inflation period to be a scalar-matter-dominated period in which the scalar field is the inflaton field, or is power-law inflation, i.e., double inflation. A simple cosmological model is assumed, as defined by
\begin{equation}
\textrm{Pre-inflation:       } a_\mathrm{p}(\eta )=b_1(-\eta-\eta_j )^r,
\end{equation}
\begin{equation}
\textrm{Slow-roll inflation:    } a_\mathrm{I}(\eta )=b_2(-\eta)^{-1-\epsilon},
\end{equation}
where
\begin{equation}
\eta_j=-(\frac{r}{1+\epsilon}+1)\eta_1 , b_1=(\frac{-1-\epsilon}{r})^r (-\eta_1)^{-1-\epsilon-r}b_2.
\end{equation}
The scale factor $a_\mathrm{I}(\eta)$ represents slow-roll inflation. Here, de-Sitter inflation ($\epsilon=0$) is not considered. Slow-roll inflation is assumed to begin at $\eta=\eta_1$. In pre-inflation, for the case of $r=2$, the scale factor $a_\mathrm{p}(\eta)$ indicates that pre-inflation is a scalar-matter-dominated period, and, for the case of $r=-q/(q-1)$, the pre-inflation is power-law inflation, where the scale factor $a_\mathrm{p}(t)\propto t^q$. 

Using above the pre-inflation model, the initial state of inflation given by Eq. (2.11) will be fixed as follows. The coefficients $c_1$ and $c_2$ are fixed using the matching condition in which the mode function and first $\eta$-derivative of the mode function are continuous at the transition time $\eta=\eta_1$. ($\eta_1$ is the time at which slow-roll inflation begins.) For simplicity pre-inflation states are assumed to be the Bunch-Davies vacuum. The coefficients $c_1$ and $c_2$ can be calculated analytically in the case of the scalar-matter-dominated period:
\begin{eqnarray}
\nonumber c_1=-\frac{i}{8z^{3/2}}\sqrt{\frac{\pi}{2}}e^{i((-1+\delta-2\epsilon)\pi/2-2z/(1+\epsilon))}
\left.\Bigl\{2z(-1-2iz-\epsilon)H^{(2)}_{\nu_2}(z)+ \right.
\end{eqnarray}
\begin{eqnarray}
\left. (4z^2+(3-2\delta+3\epsilon)(1+\epsilon+2iz))H^{(2)}_{\nu_1}(z) \right.\Bigr\},
\end{eqnarray}

\begin{eqnarray}
\nonumber c_2=-\frac{i}{8z^{3/2}}\sqrt{\frac{\pi}{2}}e^{-i((-1+\delta-2\epsilon)\pi/2+2z/(1+\epsilon))}
\left.\Bigl\{2z(-1-2iz-\epsilon)H^{(1)}_{\nu_2}(z)+ \right.
\end{eqnarray}
\begin{eqnarray}
\left. (4z^2+(3-2\delta+3\epsilon)(1+\epsilon+2iz))H^{(1)}_{\nu_1}(z) \right.\Bigr\},
\end{eqnarray}
and in the case of the double inflation model:
\begin{eqnarray}
\nonumber c_1=-\frac{\pi}{4\sqrt{q(q-1)(1+\epsilon})}e^{i\pi(6+2/(q-1)-2\delta+4\epsilon)/4}\{-qzH^{(1)}_{5/2+1/(q-1)}(zz)H^{(2)}_{\nu_1}(z) \\ 
-H^{(1)}_{3/2+1/(q-1)}(zz)(-qzH^{(2)}_{\nu_2}(z)+(1+q(-2+\delta-4\epsilon)+\epsilon)H^{(2)}_{\nu_1}(z)\}
\end{eqnarray}
\begin{eqnarray}
\nonumber c_2=-\frac{\pi}{4\sqrt{q(q-1)(1+\epsilon})}e^{i\pi(-\delta+q(1+\delta-2\epsilon)+2\epsilon)/2(q-1)}\{-qzH^{(1)}_{5/2+1/(q-1)}(zz)H^{(1)}_{\nu_1}(z) \\
-H^{(1)}_{3/2+1/(q-1)}(zz)((1+q(\delta-2-4\epsilon)+\epsilon)H^{(1)}_{\nu_1}(z)-qzH^{(1)}_{\nu_2}(z)\}
\end{eqnarray}
where $\nu_1=3/2+2\epsilon-\delta$, $\nu_2=5/2+2\epsilon-\delta$, $z=-k\eta_1$ and $zz=qz/((q-1)(1+\epsilon))$. The initial states of inflation can be written in terms of the slow-roll parameters, the start time of slow-roll inflation $\eta_1$, and the double inflation parameter $q$. Here, three slow-roll inflation models are adopted: the new inflation model with the potential term given by $V(\phi)=\lambda^2\nu^4(1-2(\phi/\nu)^p)$, where $ p=3,4$ and $\nu\approx M_\mathrm{p}$, the chaotic inflation model with the potential term given by $V(\phi)=\frac{M^4}{2(\phi/m)^a}$, where $ a=2,4,6$ and $ m\approx M_\mathrm{p}$ and the hybrid model $V(\phi)=\alpha((\nu^2-\sigma^2)^2+m^2/2 \phi^2+g^2\phi^2\sigma^4)\simeq\alpha(\nu^4+m^2/2\phi^2$), where $ \nu\approx 10^{-2}M_\mathrm{p}$ and $ m\approx 2\times10^{-5}M_\mathrm{p}$ [16]. Using the normalization value from the WMAP five-year data, we obtain the values of the spectral index and the slow-roll parameters, such as\\
New inflation: $n_\textrm{s}=0.935$, $\epsilon=1.027 \times 10^{-9}$, $\delta=-0.03228$\\
Hybrid inflation:  $n_\textrm{s}=0.9816$, $\epsilon=0.00504$, $\delta=0.000878$\\
Chaotic inflation model:\\
$\phi^2$model: $n_\textrm{s}=0.967$, $\epsilon=0.00828$, $\delta=0.000022$\\
$\phi^4$model: $n_\textrm{s}=0.950$, $\epsilon=0.01655$, $\delta=0.008298$\\
$\phi^6$model: $n_\textrm{s}=0.9334$, $\epsilon=0.0248$, $\delta=0.01657$.
 
Now, we calculate the values of the nonlinearity parameter $f^\textrm{\tiny flattened}_\textrm{\tiny NL}$. Holman and Tolley [3] showed that if the effective action for the inflaton contains the higher-derivative interaction [17] $\mathcal{L}=\sqrt{-g}\frac{\lambda}{8M^4}((\nabla\phi)^2)^2$, which is derived, for example, from k-inflation or DBI inflation, and the initial state of inflaton is not the Bunch-Davies vacuum, then the enhanced non-Gaussianity is derived as follows:
\begin{equation}
f^\textrm{\tiny flattened}_\textrm{\tiny NL}\approx \frac{\dot\phi^2}{M^4}|c_2|\big( \frac{k}{a(\eta_1)H} \big)=\frac{2\epsilon M^2_\mathrm{p}}{H^2z^3}|c_2|,
\end{equation}
where $M$ is the cutoff scale, which is the limit of effective theory, and we assum $M\approx k/a(\eta_1)$ where $\eta_1$ is the beginning time of slow-roll inflation, and $z=-k\eta_1$. The present treatment considers the effect of the length of inflation, where $z=1$ indicates that inflation starts at the time when the present-day size perturbation $k=0.002$($1/\textrm{Mpc}$) exceeds the Hubble radius in inflation (i.e., inflation of close to 60 $e$-folds). Using the values of the above parameters we can calculate the values of $|c_1|$, $|c_2|$, and $f^\textrm{\tiny flattened}_\textrm{\tiny NL}$ in terms of $z(=-k\eta_1)$. The values of $|c_2|$ change only slightly among the models, but vary with the value of $z$, as 0.0063 for $z=8$, 0.004 for $z=10$, and 0.001 for $z=20$, and $|c_1|\cong 1$. From all of the models except for the $\phi^4$ model, similar values of $f^\textrm{\tiny flattened}_\textrm{\tiny NL}$ are calculated, i.e., $f^\textrm{\tiny flattened}_\textrm{\tiny NL}\approx 120$ at $z=8$, and $f^\textrm{\tiny flattened}_\textrm{\tiny NL}\approx 40$ at $z=10$. Details are shown in Table 1. With respect to the other values of $z$, larger values of $f^\textrm{\tiny flattened}_\textrm{\tiny NL}$ can be derived at smaller $z(z<8)$, and small values of $f^\textrm{\tiny flattened}_\textrm{\tiny NL}$ can be derived at larger $z(z>20)$. Based on the above results, the value of $f^\textrm{\tiny flattened}_\textrm{\tiny NL}$ appears to depend strongly on the value of $z$, which represents the length of inflation, and the difference of the value of $f^\textrm{\tiny flattened}_\textrm{\tiny NL}$ among our three slow-roll inflation models is not large. Since the $z$-dependence of $f^\textrm{\tiny flattened}_\textrm{\tiny NL}$ is very steep, any value of $f^\textrm{\tiny flattened}_\textrm{\tiny NL}$ can be derived at some point of $z$. We next consider the case of double inflation, the value of $f^\textrm{\tiny flattened}_\textrm{\tiny NL}$ is 100 at $3<z<4$ in the chaotic inflation, at $4<z<5$ in the case of new inflation, and at $z\approx 3$ in the case of hybrid inflation. With respect to the $q$-dependence ( $a(t)\propto t^q$), the values of $f^\textrm{\tiny flattened}_\textrm{\tiny NL}$ are similar at very large $q$ but change at $q\approx 100$. The details are shown in Tables 2-4.

\section{Discussion}
We have derived a new property of the proposed finite inflation model. The possibility of large non-Gaussianity is demonstrated. The proposed inflation model is a finite length inflation model with an effective higher derivative interaction, where slow-roll inflation is adopted as inflation and a scalar-matter-dominated period or power inflation is adopted as pre-inflation. Owing to the existence of pre-inflation, the initial state in inflation is not the Bunch-Davies state, but is instead a more general state. The coefficients $c_1$ and $c_2$ can be analytically calculated. Using Holman and Tolley's formula of the nonlinearity parameter $f^\textrm{\tiny flattened}_\textrm{\tiny NL}$, we calculated the value of $f^\textrm{\tiny flattened}_\textrm{\tiny NL}$. For the case in which the scalar-matter-dominated period is considered to be pre-inflation, large values of $f^\textrm{\tiny flattened}_\textrm{\tiny NL}(f^\textrm{\tiny flattened}_\textrm{\tiny NL}\approx 100)$ are obtained at $8<z<10$ in all the models considered herein, and similar results are derived for the case of double inflation at $3<z<4$. These ranges can be written as 60-63 $e$-folds. This length is similar to that obtained when the suppression of CMB angular power spectrum of $l=2$ was derived using the inflation models described in previous papers [9], although such spectral suppression is not inconsistent when considering cosmic variance. On the experimental value of $f^\textrm{\tiny flattened}_\textrm{\tiny NL}$ , the orthogonal shape ($f^\textrm{\tiny orthog}_\textrm{\tiny NL}$) is peaked both on equilateral-triangle configurations ($f^\textrm{\tiny equil}_\textrm{\tiny NL}$) and on flattened-triangle configurations ($f^\textrm{\tiny flattened}_\textrm{\tiny NL}$) [18], but we think we need further considerations to derive the constraint of $f^\textrm{\tiny flattened}_\textrm{\tiny NL}$ from the constraints of $f^\textrm{\tiny orthog}_\textrm{\tiny NL}$ and $f^\textrm{\tiny equil}_\textrm{\tiny NL}$. Therefore, we do not show it here.  We assume such a high-derivative interaction in order to obtain non-linearity and effective interactions for slow-roll interaction. This high-derivative interaction appears to influence the parameters of slow-roll inflation. In order to clarify this problem, we must investigate a concrete inflation model such as k-inflation or DBI inflation. In the future, we would like to apply the proposed method to other inflation models and investigate the dependence of the length of inflation on $f^\textrm{\tiny flattened}_\textrm{\tiny NL}$.

\acknowledgments
The authors would like to thank the staff of Osaka Electro-Communication University for their valuable discussions.

\begin{table}[ht]
\caption{Values of $f^\textrm{\tiny flattened}_\textrm{\tiny NL}$ for the case of the matter-dominated period as pre-inflation}
\begin{center}
\begin{tabular}{|r|rr|r|rrr|} \hline
& \multicolumn{2}{|c|}{New inflation} & Hybrid & \multicolumn{3}{|c|}{Chaotic inflation} \\
     &  $p=3$  &  $p=4$  &       & $\phi^2$ & $\phi^4$ & $\phi^6$ \\
\hline
$z=8$  & 123.8 & 123.7 & 122.7 & 123.8 & 187.7 & 126.4 \\
\hline
$z=10$ & 40.5 &  40.4 &  40.1 & 40.4  &  61.3 &  41.3 \\
\hline
$z=20$ & 1.26 &  1.26 &  1.25 & 1.26  &  1.91 &  1.28 \\
\hline
\end{tabular}
\end{center}
\end{table}

\begin{table}[ht]
\caption{Values of $f^\textrm{\tiny flattened}_\textrm{\tiny NL}$ in the hybrid inflation for double inflation}
\begin{center}
\begin{tabular}{|r|r|r|r|r|r|} \hline
	&$q=10^5$	&$q=10^4$	&$q=10^3$	&$q=10^2$	&$q=10$ \\ \hline
$z=3$	&108.5	&109.1	&115.3	&190.6	&1096.5 \\ \hline
$z=4$	&23.4	&23.6	&25.3	&45.1	&266.8 \\ \hline
$z=5$	&7.24	&7.30	&7.93	&14.8	&88.6 \\ \hline
\end{tabular}
\end{center}
\end{table}

\begin{table}[ht]
\caption{Values of for $f^\textrm{\tiny flattened}_\textrm{\tiny NL}$ the new inflation case of $n=3$ and for the new inflation case of $n=4$ for double inflation}
\begin{center}
$n=3$ \\[2mm]
\begin{tabular}{|r|r|r|r|r|r|} \hline
	&$q=10^5$	&$q=10^4$	&$q=10^3$	&$q=10^2$	&$q=10$ \\ \hline
$z=4$	&254.0	&254.1	&256.0	&275.1	&478.5 \\ \hline
$z=5$	&81.8	&81.9	&82.5	&89.1	&157.7 \\ \hline
$z=6$	&32.5	&32.6	&32.8	&35.6	&63.6 \\ \hline
\end{tabular}
\\[5mm]
$n=4$ \\[2mm]
\begin{tabular}{|r|r|r|r|r|r|} \hline
	&$q=10^5$	&$q=10^4$	&$q=10^3$	&$q=10^2$	&$q=10$ \\ \hline
$z=4$	&194.9	&195.1	&197.0	&216.5	&424.2 \\ \hline
$z=5$	&62.8	&62.9	&63.5	&70.2	&140.1 \\ \hline
$z=6$	&25.0 	&25.0	&25.3	&28.0	&56.5 \\ \hline
\end{tabular}
\end{center}
\end{table}

\begin{table}[ht]
\caption{Values of $f^\textrm{\tiny flattened}_\textrm{\tiny NL}$ for the Chaotic inflation in the cases $\phi^2$,  $\phi^4$, and $\phi^6$ for double inflation}
\begin{center}
$\phi^2$ model \\[2mm]
\begin{tabular}{|r|r|r|r|r|r|} \hline
	&$q=10^5$	&$q=10^4$	&$q=10^3$	&$q=10^2$	&$q=10$ \\ \hline
$z=3$	&227.6	&228.2	&234.7	&306.9	&1196.5 \\ \hline
$z=3.5$	&100.6	&100.9	&104.2	&140.0	&561.2 \\ \hline
$z=4$	&49.8	&50.0	&51.8	&71.1	&291.0 \\ \hline
\end{tabular}
\\[5mm]
$\phi^4$ model \\[2mm]
\begin{tabular}{|r|r|r|r|r|r|} \hline
	&$q=10^5$	&$q=10^4$	&$q=10^3$	&$q=10^2$	&$q=10$ \\ \hline
$z=3$	&181.1	&181.5	&185.6	&242.9	&1130.3 \\ \hline
$z=3.5$	&76.1	&76.4	&78.6	&108.1	&530.0 \\ \hline
$z=4$	&36.1	&36.2	&37.5	&53.9	&274.4 \\ \hline
\end{tabular}
\\[5mm]
$\phi^6$ model \\[2mm]
\begin{tabular}{|r|r|r|r|r|r|} \hline
	&$q=10^5$	&$q=10^4$	&$q=10^3$	&$q=10^2$	&$q=10$ \\ \hline
$z=3$	&165.5	&165.5	&165.4	&191.2	&1061.5 \\ \hline
$z=3.5$	&67.1	&67.1	&67.1	&81.4	&497.6 \\ \hline
$z=4$	&30.6	&30.6	&30.6	&39.0	&257.6 \\ \hline
\end{tabular}
\end{center}
\end{table}

\end{document}